\documentclass[twocolumn]{aastex62}
\usepackage{graphics,epsf}
\usepackage{amsmath}                
\usepackage{amsfonts}               
\usepackage{amssymb}                
\usepackage{epsfig}                 
\usepackage{graphicx}
\usepackage{float}
\usepackage{color}
\usepackage[para,online,flushleft]{threeparttable}

\newcommand{\cm}{{~\rm cm}}
\newcommand{\km}{{~\rm km}}
\newcommand{\s}{{~\rm s}}

\newcommand{\g}{{~\rm g}}

\newcommand{\erg}{{~\rm erg}}
\newcommand{\yr}{{~\rm yr}}

\newcommand{\days}{{~\rm days}}

\begin{document}

\title{Explaining recently studied intermediate luminosity optical transients (ILOTs) with jet powering}

\author[0000-0003-0375-8987]{Noam Soker}
\affiliation{Department of Physics, Technion, Haifa, 3200003, Israel;  soker@physics.technion.ac.il}
\affiliation{Guangdong Technion Israel Institute of Technology, Shantou 515069, Guangdong Province, China}

\author{Noa Kaplan}
\affiliation{Department of Physics, Technion, Haifa, 3200003, Israel;  soker@physics.technion.ac.il}

\begin{abstract}
We apply the jet-powered ILOT scenario to two recently studied intermediate luminosity optical transients (ILOTs), and find the relevant shell mass and jets' energy that might account for the outbursts of these ILOTs. In the jet-powered ILOT scenario accretion disk around one of the stars of a binary system launches jets. The interaction of the jets with a previously ejected slow shell converts kinetic energy to thermal energy, part of which is radiated away. We apply two models of the jet-powered ILOT scenario. In the spherical shell model the jets accelerate a spherical shell, while in the cocoon toy model the jets penetrate into the shell and inflate hot bubbles, the cocoons. We find consistent results. For the ILOT (ILRT: intermediate luminosity red transient) SNhunt120 we find the shell mass and jets' energy to be $M_{\rm s} \simeq 0.5-1 M_\odot$ and $E_{\rm 2j} \simeq 5 \times 10^{47} \erg$, respectively. The jets' half opening angle is $\alpha_j \simeq 30^\circ - 60 ^\circ$. For the second peak of the ILOT (luminous red nova) AT~2014ej we find these quantities to be $M_{\rm s} \simeq 1-2 M_\odot$ and $E_{\rm 2j} \simeq 1.5 \times 10^{48} \erg$, with $\alpha_j \simeq 20^\circ - 30 ^\circ$. The models cannot tell whether these ILOTs were powered by a stellar merger that leaves one star, or by mass transfer where both stars survived.  
In both cases the masses of the shells and energies of the jets suggest that the binary progenitor system was massive, with a combined mass of $M_1 + M_2 \ga 10 M_\odot$.  
\end{abstract}


\section{Introduction}
\label{sec:intro}

The transient events with peak luminosities above those of classical novae and below those of typical supernovae  might differ from each other by one or more properties, like the number of peaks in the light curve, total power, progenitor masses, and powering mechanism (e.g. \citealt{Mouldetal1990, Bondetal2003, Rau2007, Ofek2008, Masonetal2010, Kasliwal2011, Tylendaetal2013, Kasliwaletal2012, Blagorodnovaetal2017, Kaminskietal2018, Pastorelloetal2018, BoianGroh2019, Caietal2019, Jencsonetal2019, Kashietal2019, PastorelloMasonetal2019, Howittetal2020, Jones2020, Klenckietal2020}).
They form a heterogeneous group of `gap transients'. 

We study those transients that are powered by an accretion process that releases gravitational energy.
The accretion process might be a mass transfer from one star to another, or an extreme case of stellar merger, where either one star destroys another, or one star (or a planet; \citealt{RetterMarom2003, Bearetal2011, KashiSoker2017, Kashietal2019}) enters the envelope of a larger star to start a common envelope evolution (e.g., \citealt{Tylendaetal2011, Ivanovaetal2013a, Nandezetal2014, Kaminskietal2015, MacLeodetal2017, Segevetal2019, Schrderetal2020, MacLeodLoeb2020, Soker2020Galaxy}). 
We refer to all these systems as intermediate luminosity optical transients (ILOTs). 

In cases where both stars survive and stay detached the binary system can experience more than one outburst, and can have several separated peaks in its light curve. This is the case for example in the grazing envelope evolution (\citealt{Soker2016GEEI}). 
The same holds when the binary system forms a temporary common envelope. Namely, the more compact companion enters the envelope and then gets out. An example of the later process is the repeating common envelope jets supernova (CEJSN) impostor scenario \citep{Gilkisetal2019}. In a CEJSN impostor event a neutron star (or a black hole) gets into the envelope of a giant massive star, accretes mass and launches jets that power an ILOT event (that might be classified as a supernova impostor), and then gets out of the envelope \citep{SokerGilkis2018, Gilkisetal2019, YalinewichMatzner2019}. 
 
Mass outflow accompanies the bright outbursts of ILOTs. Many studies attribute the powering of ILOTs, both the kinetic energy of the outflow and the radiation, to stellar binary interaction processes (e.g., \citealt{SokerTylenda2003, TylendaSoker2006AA, Kashietal2010, McleySoker2014, Pejchaetal2016a, Pejchaetal2016b, Soker2016GEEI, MacLeodetal2018, Michaelisetal2018, PastorelloMasonetal2019}). As a fast outflow hits a previously ejected slower outflow, the collision channels kinetic energy to radiation. There are two types of binary scenarios in that respect, those that take the main collision to take place in and near the equatorial plane (e.g., \citealt{Pejchaetal2016a, Pejchaetal2016b, MetzgerPejcha2017, HubovaPejcha2019}), and those that attribute the main collision to fast polar outflow, i.e., jets.     
In most of the cases with high mass accretion rates that power ILOTs, the high-accretion-powered ILOT (HAPI) model \citep{KashiSoker2016, SokerKashi2016TwoI}, the accretion of mass is likely to be through an accretion disk. 
This accretion disk is very likely to launch two opposite jets. If the jets collide with a previously ejected slow shell an efficient conversion of kinetic energy to radiation might take place. This is the \textit{jet-powered ILOT scenario}. 
 
In a recent study  \cite{Soker2020ILOT} argues that the jets-shell interaction of the jet-powered ILOT scenario is more efficient in converting kinetic energy to radiation than collision of equatorial ejecta. He further applies a simple jet-shell interaction model to three ILOTs, the Great Eruption of Eta Carinae \citep{DavidsonHumphreys1997}, which is a luminous blue variable (LBV), to V838~Mon \citep{Munarietal2002} that is a stellar merger (also termed luminous red nova; LRN), and to the ILOT V4332~Sgr that has a bipolar structure \citep{Kaminskietal2018}. 
We apply this simple spherical shell model to two other ILOTs (Sections \ref{subsec:SNhunt120} and \ref{subsec:AT2014ej}). 

As said, in this study we use the term ILOT \citep{Berger2009, KashiSoker2016, MuthukrishnaetalM2019}. There are different classifications of the heterogeneous class of transients, like the one by \cite{KashiSoker2016}\footnote{See \url{http://physics.technion.ac.il/~ILOT/} for an updated list.},  the one by \cite{PastorelloMasonetal2019} and \cite{PastorelloFraser2019}, and also by \citet{Jencsonetal2019}. 
Some refer to transients from stellar merger by LRNe and to outbursts that involve a massive giant star by intermediate luminosity red transients (ILRTs). 
We simply refer to all transients that are powered by gravitational energy of mass transfer (or merger), the HAPI model, as ILOTs. This saves us the need to classify a specific event by its unknown progenitors. We are mainly interested in the roles of jets, that might play a role in all types of ILOTs (although not in all ILOTs). 

Two recent studies of two ILOTs support two crucial  ingredients of the jet-powered ILOT scenario. 
\cite{Blagorodnovaetal2020} study the ILOT  M31-LRN-2015 that is possibly a merger remnant (some earlier studies related to this ILOT include \citealt{Williamsetal2015, Lipunovetal2017, MacLeodetal2017, MetzgerPejcha2017}). \cite{Blagorodnovaetal2020} estimate the primary mass to be $M_1 \simeq 5 M_\odot$ and deduce that during the two years pre-outburst activity the system lost a mass of about $>0.14 M_\odot$. Such a pre-outburst formation of a shell (circumbinary matter) is an important ingredient in the jet-powered ILOT scenario. 
 
In other recent papers \cite{Kaminskietal2020Nova1670, Kaminskietal2021Nova1670} study in details the ILOT (stellar-merger candidate) Nova~1670 (CK~Vulpeculae). This 350-years old nebula has a bipolar structure \citep{Sharaetal1985} with an `S' shape along the long axis \citep{Kaminskietal2020Nova1670, Kaminskietal2021Nova1670}. This is an extremely strong indication of shaping by precessing jets. We take it to imply that the jet-powered ILOT scenario accounts for Nova~1670. 
The intervals from the first to second peak and from the second to third peak in the triple-peaks light curve are about equal at about 1 year \citep{Sharaetal1985}. We take it to imply a multiple jets-launching episodes, or more likely in this case, a variability in jets' launching power as the jets precess. 

These two recent studies, and in particular the clear demonstration of an `S' shape morphology of the ILOT Nova~1670 \citep{Kaminskietal2020Nova1670, Kaminskietal2021Nova1670} motivate us to apply the jet-powered ILOT scenario to two recently studied ILOTs. 
We emphasise that our main aim is to find plausible parameters for these two recently studied ILOTs in the frame of the jet-driven model, as the formation of jets in binary merger can be very common (e.g., \citealt{LopezCamaraetal2020} and references therein).
In section \ref{sec:Features} we describe the basic features of the jet-powered ILOT scenario and apply it in a simple way to the ILOTs SNhunt120 and AT~2014ej. In section \ref{sec:BipolarToyModel} we build a more sophisticated toy model to describe the jet-powered ILOT scenario and apply it to these two ILOTs. 
We summarise in section \ref{sec:summary}.

\section{The jet-powered ILOT scenario} 
\label{sec:Features}
\subsection{Features of the spherical shell model} 
\label{subsec:features}
The basic flow structure of the jet-powered ILOT scenario is as follows \citep{Soker2020ILOT}. A binary interaction leads to the ejection of a shell, spherical or not, at velocities of tens to hundreds of $\km \s^{-1}$. The shell ejection period can last from few weeks to several years. In a delay of about days to several months (or even a few years) the binary system launches two opposite jets. The jets collide with the shell, an interaction that converts kinetic energy, mainly of the jets, to radiation. 
      
There are two types of evolutionary channels to launch jets. (1) The more compact secondary star accretes mass from the primary star and launches the jets, as in the jet-powered ILOT scenario of the Great Eruption of Eta Carinae (e.g., \citealt{Soker2007, KashiSoker2010a}). The binary stellar system might stay detached, might experience the grazing envelope evolution, and/or enters a common envelope evolution. In this case the binary systems might experience several  jets-launching episodes.  
(2) The primary star gravitationally destroys the secondary star to form an accretion disk around the primary star, and this accretion disk launches the jets. In this case there is one jets-launching episode, although the jets' intensity can very with time. 

\cite{Soker2020ILOT} obtains the following approximate relations for jets that interact with a slower spherically symmetric shell and power an ILOT. We refer to this model as the spherical shell model. \cite{Soker2020ILOT} considers jets-shell interaction that (1) transfers a large fraction of the kinetic energy of the outflow to radiation, and (2) radiates much more energy than what recombination of the outflowing gas can supply. 
\cite{Soker2020ILOT} considers two opposite fast jets that hit a uniform spherical shell and \textit{accelerate the entire shell.} In section \ref{sec:BipolarToyModel} we build a toy model where the jets penetrate into the shell and interact with shell's material only in the polar directions. 

In the simple flow structure that \cite{Soker2020ILOT} considers the relevant properties of the jets are their half opening angle $\alpha_j \ga 10^\circ$, velocity $v_j \approx 10^3 \km \s^{-1}$, and their total mass $M_{\rm 2j} \approx 0.01-1 M_\odot$. With a conversion efficiency of jets' kinetic energy to radiation $f_{\rm rad}$, the total energy in radiation is 
\begin{equation}
E_{\rm rad, j}  = 10^{48} f_{\rm rad}  
\left(\frac {M_{\rm 2j}}{0.1 M_\odot} \right) 
\left(\frac {v_{\rm j}}{1000 \km \s^{-1}} \right)^2 \erg .  
\label{eq:Eradj}
\end{equation}
The relevant properties of the spherical shell are its velocity $v_{\rm s} \ll v_{\rm j}$, mass $M_{\rm s}$, radius $r_{\rm s}$, and width $\Delta r_{\rm s}$.

The jet-shell interaction converts kinetic energy, mainly of the jets, to thermal energy. The hot bubbles that the jets inflate lose their energy adiabatically by accelerating the shell and non-adiabatically by radiation. The adiabatic cooling proceeds on a typical time scale that is the expansion time $t_{\rm exp}$, while energy losses to radiation occurs during a typical photon-diffusion time out $t_{\rm diff}$. 
Namely, the relative rates, $\dot E/E$, of adiabatic and radiative energy losses are $t^{-1}_{\rm exp}$ and $t^{-1}_{\rm diff}$, respectively. This implies that the fraction of energy that ends in radiation is 
\begin{equation}
f_{\rm rad} \simeq \frac {t^{-1}_{\rm diff}}{t^{-1}_{\rm diff} + t^{-1}_{\rm exp}}= \left( 1+ \frac{t_{\rm diff}}{t_{\rm exp}} \right)^{-1}.
\label{eq:frad}
\end{equation}
For the simple spherically symmetric geometry he assumes, \cite{Soker2020ILOT} estimates the two time scales to be 
\begin{eqnarray}
\begin{aligned} 
t_{\rm exp} \approx & 73    
\left( \frac{ r_{\rm s}}{10^{14} \cm} \right)  
\left( \frac{v_{\rm j}}{1000 \km \s^{-1}} \right)^{-1} 
\\ & \times 
\left[ \frac{M_{\rm 2j}}{0.1(M_{\rm 2j}+M_{\rm s})} \right]^{-1/2}
\days,   
\end{aligned}
\label{eq:texp}
\end{eqnarray}
and 
\begin{eqnarray}
\begin{aligned} 
t_{\rm diff} & \simeq \frac{3 \tau \Delta r_{\rm s}}{c} 
\simeq  55   
\left( \frac{M_{\rm s}}{1 M_\odot} \right)
\left( \frac{\kappa}{0.1 \cm^2 \g^{-1}} \right) 
\\ & \times 
\left( \frac{r_{\rm s}}{10^{14} \cm} \right)^{-1}
\left( \frac{\Delta r_{\rm s}}{0.3 r_{\rm s}} \right) \days,
\end{aligned}
\label{eq:tdiff}
\end{eqnarray}
where $\tau=\rho_{\rm s} \kappa \Delta r_{\rm s}$ is the optical depth of the shell, $\kappa$ is the opacity, and $c$ is the light speed.  
The relevant ratio to substitute in equation (\ref{eq:frad}) is 
\begin{eqnarray}
\begin{aligned} 
\frac { t_{\rm diff}}{t_{\rm exp}}  & \approx 0.75  %
\left( \frac{M_{\rm s}}{1 M_\odot} \right)
\left( \frac{\kappa}{0.1 \cm^2 \g^{-1}} \right) 
\left( \frac{v_{\rm j}}{1000 \km \s^{-1}} \right)
\\ & \times 
\left( \frac{r_{\rm s}}{10^{14} \cm} \right)^{-2}
\left( \frac{\Delta r_{\rm s}}{ 0.3 r_{\rm s}} \right) 
\left[ \frac{M_{\rm 2j}}{0.1(M_{\rm 2j}+M_{\rm s})} \right]^{1/2}.
\end{aligned}
\label{eq:tdiffdexp}
\end{eqnarray}
We emphasise that we do not assume any value for the jets' energy $E_{\rm 2j}$. We rather take the jets' velocity from observations, and use the time scale of the ILOT together with an assumed opacity to find the mass in the shell (equation \ref{eq:tdiff}). We then calculate the efficiency $f_{\rm rad}$ together with the mass in the jets (or their energy) to fit the total radiated energy (equations \ref{eq:frad} and \ref{eq:tdiffdexp}).   
    
\cite{Soker2020ILOT} applies this spherical shell model of the jet-powered ILOT radiation to the ILOT (LRN) V838 Mon, to the Great Eruption of Eta Carinae which is an LBV, and to the ILOT V4332~Sgr.
He could find plausible set of shell and  jets parameters that might explain these ILOTs. 
Here we apply the spherical shell model to the ILOT (ILRT) SNhunt120 and to the ILOT (LRN) AT~2014ej. 
We summarise the plausible physical parameters of the ILOT events in Table \ref{Table:Parameters}, and explain their derivation in sections \ref{subsec:SNhunt120} and \ref{subsec:AT2014ej}. We emphasise that due to the very simple model we apply here, e.g., we use a spherical shell and we keep the opacity and shell thickness constant, the properties of the jets and shells we derive are very crude, and might even not be unique. Nonetheless, they demonstrate the potential of the jet-powered ILOT scenario to account for many ILOTs. 
The opacity of a fully ionised gas that is appropriate for ILOTs is $\kappa \simeq 0.3 \cm^2 \g^{-1}$ (e.g., \citealt{Ivanovaetal2013a, SokerKashi2016TwoI}). We expect that in the outer parts of the shell hydrogen is partially neutral, and that opacity is therefore lower. Therefore, we scale with $\kappa=0.1 \cm^2 \g^{-1}$.
\begin{table}
\footnotesize
\centering
\begin{tabular}{|l|c|c|c|}
\hline
Property  & SNhunt120 & AT~2014ej     & AT 2014ej      \\
          &           & $1^{\rm st}$p & $2^{\rm nd}$p \\ 
 \hline
 \hline  
[O] Radiated energy &   $4 \times 10^{47}$ & $10^{48}$ & $1.4 \times 10^{48}$\\
$E_{\rm rad}$ (erg) &                      & (assumed) & \\
   \hline
[O] Time scale      &  10-20                    & $\approx 20 $ & 40 \\
(days)                &                      &  & \\  
     \hline
[O] Photosphere   & $2\times 10^{14}$         & $2.5\times 10^{14}$ & $2.5\times 10^{14}$  \\ 
$R_{\rm BB}$ (cm)& &  & \\   
   \hline 
   \hline    
[J] Shell mass        & 0.7  &  1.5  & 1.5 \\
$M_{\rm s} (M_\odot)$  &     &        &  \\
   \hline 
[J] Jets' mass        & 0.045  &  0.1  &  0.15\\
$M_{\rm 2j} (M_\odot)$ &     &    &  \\
 \hline
[J] Jets' Energy   & $4.5\times 10^{47}$&$1.1\times 10^{48}$ &  $1.6\times 10^{48}$\\
$E_{\rm 2j}$ (erg) &     &    &  \\
\hline
[J] Emission efficiency  & 0.9  & 0.9   &  0.9 \\
$f_{\rm rad}$         &   &     &     \\
 \hline
\end{tabular}
\caption{Summary of plausible approximate  values of parameters in the spherical-shell ILOT model of \cite{Soker2020ILOT} for the ILOTs SNhunt120 and AT~2014ej. We assume that AT~2014ej was powered by two jet-launching episodes, each accounting for one of the two peaks in the light curve. The symbol `[O]' in the first column implies a quantity we take from observations, while `[J]' indicates that we derive the plausible parameter. We derive these parameters under the assumption of a constant opacity of $\kappa=0.1 \cm^2 \g^{-1}$ and a constant shell width of $\Delta r_{\rm s}=0.3r_{\rm s}$. In both ILOTs observation suggest jets' velocity of $v_{\rm j} \simeq 1000 \km s^{-1}$ which we also use here. }
\label{Table:Parameters}
\end{table}
    
\subsection{The ILOT SNhunt120} 
\label{subsec:SNhunt120}
\cite{Stritzingeretal2020SNhunt120} study the ILOT (ILRT) SNhunt120 and find the following relevant properties. The velocities of different emission lines are in the range of $\simeq 300-1800 \km \s^{-1}$, with a typical velocity of $\approx 10^3 \km \s^{-1}$. The typical photospheric radius is $R_{\rm BB} \simeq 2 \times 10^{14} \cm$. The time to double the luminosity at rise is about 10 days, and the decline time to half the maximum luminosity is about 20 days. The total energy in radiation is $E_{\rm rad} \simeq 4 \times 10^{47} \erg$.
\cite{Stritzingeretal2020SNhunt120} further find that existing electron capture supernova models over-predict the energy in radiation. We do not consider this event to be a supernova, but rather an ILOT. 

Following these parameters we scale the parameters for SNhunt120 with  $v_j \simeq 1000 \km \s^{-1}$ and $r_s \simeq 2 \times 10^{14} \cm$. 
To get a photon diffusion time of about the decline time of 20 days, we find from equation (\ref{eq:tdiff}) for $\kappa=0.1 \cm^2 \g^{-1}$ and $\Delta r_{\rm s} =0.3 r_{\rm s}$ that $M_{\rm s } \approx 0.7 M_\odot$. 
For an opacity of $\kappa=0.3 \cm^2 \g^{-1}$ and somewhat a thicker shell with $\Delta r_{\rm s} = 0.5 R_{\rm s}$, the required shell mass is only $M_{\rm s } \approx 0.15 M_\odot$

Equation (\ref{eq:tdiffdexp}) gives then $t_{\rm diff}/t_{\rm exp} \approx 0.1$, and from equation (\ref{eq:frad}) $f_{\rm rad} \simeq 0.9$. To account for the emitted energy, we find from equation (\ref{eq:Eradj}) that the mass in the two jets is $M_{\rm 2j} \approx 0.045 M_\odot (v_{\rm j} / 1000 \km \s^{-1})^{-2}$.   
    
In this analysis there is no reference to the shell velocity, beside that it should be much lower than the jets' velocity. This implies here
$100 \km \s^{-1} \la v_s \la 500 \km \s^{-1}$. To reach a distance of $r_{\rm s}=2 \times 10^{14}$ the binary system ejected the shell about $\Delta t_{\rm s} \simeq 0.6 (v_s/100 \km \s^{-1})^{-1} \yr$ before detection. 
The kinetic energy in the shell for these parameters of $M_{s} \simeq 0.7 M_\odot$ and $v_{\rm s} \simeq 100 \km \s^{-1}$ is about $15 \%$ of the jets' energy. In any case, most of the kientic enrgy of the shell does not convert to radiation. 

In case that the secondary star launches the jets with a mass of $M_{\rm 2j} \simeq 0.045M_\odot$, it should accrete a mass of $M_{\rm acc,2} \simeq 0.45 M_\odot$ from a more evolved primary star, possibly a giant. This implies that the secondary star should be a massive star itself. We are therefore considering a massive binary system. Alternatively, it is possible that the primary star destroyed the secondary star of mass $M_2 \simeq 0.3-1 M_\odot$ to form an accretion disk that launched the jets. The primary is then a massive main sequence star, and the secondary is not yet settled on the main sequence, such that its average density is lower than that of the primary star (as in the merger model of V838~Mon; \citealt{TylendaSoker2006AA}). In any case, the primary star mass can be in the range of $M_1 \approx 10 M_\odot$, similar to the range that \cite{Stritzingeretal2020SNhunt120} consider. Since there is only one jets-launching episode, the jet-powered ILOT scenario does not directly refer in the case of SNhunt120 to the question of which of these two evolutionary routes apply here. 

\subsection{The ILOT AT~2014ej} 
\label{subsec:AT2014ej}

\cite{Stritzingeretal2020AT2014ej} study the ILOT (LRN) AT~2014ej. They find that the light curve of models of equatorial collision  (\citealt{MetzgerPejcha2017}; section \ref{sec:intro}), under-predict the  luminosity. We therefore consider powering by jets, i.e., polar collision. 

\cite{Stritzingeretal2020AT2014ej} find that AT~2014ej has slow component(s) moving at $\approx 100 \km \s^{-1}$ and fast component(s) moving at $\approx 1000 \km \s^{-1}$. The total radiated energy is $E_{\rm rad} \approx 2 \times 10^{48} \erg$, with two large peaks in the light curve. From discovery to first minimum 20 days later, the luminosity decreased from  $L_0=3.2 \times 10^{41} \erg \s^{-1}$ to $L_{\rm min,1} = 1.2 \times 10^{41} \erg \s^{-1}$. Over the next 35 days the luminosity increased to $L_{\rm AT} \equiv L_{\rm peak,2} \simeq 2.6 \times 10^{41} \erg \s^{-1}$, after which the luminosity decreased over a time scale of several weeks. The photosphere was hotter in the first peak than in the second one. 
The photosphere (black body surface) moderately followed the behavior of the luminosity, and first declined somewhat and then increased somewhat. Its approximate average value is $R_{\rm BB} \simeq 2.5 \times 10^{14} \cm$.
 
In the jet-powered ILOT scenario such multiple-peaks can be accounted for by multiple jet-launching episodes.
From \cite{Stritzingeretal2020AT2014ej} we find that the radiated energy from detection to first minimum (0 to 20 days) is $\simeq 4 \times 10^{47} \erg$. If we take a similar energy at rise, the energy in the first peak is $E_{\rm rad,1p} \approx 10^{48} \erg$. The energy in the second peak, from 20 to about 95 days, is $E_{\rm rad,2p} \approx 1.4 \times 10^{48} \erg$. The outburst of V838~Mon has a similar qualitative behavior with three peaks and three declines in the photospheric radius \citep{Tylenda2005}.

In AT~2014ej the two peaks have about the same energy (under our assumption), but the second peak is slower by a factor of about two. From equation (\ref{eq:tdiff}) the mass in the shell should be larger in the second peak by a factor of two, $\simeq 2M_\odot$ instead of $\simeq 1 M_\odot$. We do not expect the system to lose much more slow mass in that short time. The difference in the time scales of the two peaks might come from different values of $\kappa$ and/or $\Delta r_{\rm s}$ between the two peaks, rather than from different shell masses. This can also be accompanied by precessing jets, i.e., the jets' axes in the two jet-launching episodes have different directions.  In the present study we use a simple model and do not calculate the opacity, and so we simply take for both peaks $M_{\rm s} \simeq 1.5 M_\odot$. 

From the equations of section \ref{subsec:features} we derive the crude plausible values of the shell mass, jets' energy, and emission efficiency for the two peaks, as we list in Table \ref{Table:Parameters}.  

According to the jet-powered ILOT scenario the two distinguished peaks result from two jets-launching episode. This suggests that the secondary star, possibly in an eccentric orbit, accreted mass and launches the jets. Most likely, the secondary star survived the interaction.  

\section{A bipolar toy-model} 
\label{sec:BipolarToyModel}
\subsection{The cocoon toy-model} 
\label{subsec:interaction_TM}

In the simple spherical-shell model that we apply in section \ref{sec:Features} the jets interact with the entire shell \citep{Soker2020ILOT}. 
We now turn to a more realistic toy model where the jets interact only with the shell segments along the polar directions. 
In this `cocoon toy model' the jet-shell interaction inflates a `cocoon', i.e., a relatively hot bubble composed of the post-shock shell material and post-shock jet's material. We further simplify the interaction by assuming that the jets' activity time period is short, such that we can treat the jet-shell interaction that creates the cocoon as a `mini explosion'. We base the cocoon toy model on our usage of this model to account for peaks in the light curves of core collapse supernovae (\citealt{KaplanSoker2020a}; for the geometry of a jet-ejecta interaction in core collapse supernova see the three-dimensional simulations of \citealt{AkashiSoker2020}). 
In the cocoon toy model we only calculate the timescale of the emission peak (eruption) and its maximum luminosity (or total energy). We do not calculate the shape of the light curve, but rather assume a simple shape for the light curve. We then calculate the total radiated energy by integrating the luminosity over time.  
  
We assume that each mini-explosion that results from jet-shell interaction is spherically symmetric around the jet-shell interaction point \citep{AkashiSoker2020}, and that cooling is due to photon diffusion and adiabatic expansion. These assumptions allow us to determine the luminosity and the time scale of each mini-explosion. 
As we deal with ILOTs where the total radiated energy is larger than the recombination energy of the outflowing gas, we neglect the recombination energy.
Like \cite{KaplanSoker2020a} we use equations (4) from \cite{KasenWoosley2009} to calculate the time of maximum luminosity $t_{\rm j}$ and the maximum luminosity $L_{\rm j}$ for one jet. These expressions read
\begin{eqnarray}
\begin{aligned} 
& t_{\rm j} = \left( \frac{3}{2^{5/2} \pi^2 c} \right)^{1/2} E_{\rm j}^{-1/4} M_{\rm js}^{3/4} \kappa_{\rm c}^{1/2},\\ 
& L_{\rm j} = \frac{2 \pi c}{3} \sin{\alpha_{\rm j}}
\: \beta \:
M_{\rm js}^{-1} E_{\rm j} \kappa_{\rm c}^{-1} R_{\rm BB},
\end{aligned}
\label{eq:jet}
\end{eqnarray}
where $E_{\rm j}$, $M_{\rm js}$, $\kappa_{\rm c}$, $\alpha_{\rm j}$, $\beta$ and $R_{\rm BB}$ are the energy that one jet deposits into the shell, the mass in the interaction region of one jet with the shell, the opacity in the cocoon, the half opening angle of the jet, the distance of the jet-ejecta interaction relative to the shell's outer edge (the photosphere radius $R_{\rm BB}$), and the photospheric radius of the shell, respectively. 
Namely, in this model the mini explosion takes place at a radius (measured from the center of the binary system) of $r_{\rm me} = \beta R_{\rm BB}$. There is one mini-explosion on each of the two polar regions.  The value of $r_{\rm me}$ is constant and does not change with time. What increases with time is the radius of the cocoon itself, $a_{\rm c}$, that is measured from the place of the mini explosion.   

We build the light curve of the jet as follows. We assume that the shape of the rise of the peak to maximum luminosity is similar to the rise to maximum of the light curve of a core collapse supernova (based on photometric data of SN 2008ax, taken from The Open Supernova Catalog \citealt{Guillochonetal2017}). Since the light curve of the jet does not have a tail powered by radioactive processes and recombination, we take the decline of the mini-explosion from maximum to be symmetric to its rise. Again, we do not try to fit the light curve. We rather only derive the properties of the jets that might lead to an event that has the same timescale, luminosity and radiated energy. We assume a light curve, but our results are not sensitive to the exact shape of the light curve we assume.  

We turn to estimate the jets' properties that according to the cocoon toy model might fit the eruption times and luminosities of the ILOTs SNhunt120 (section \ref{subsec:fit_SNhunt120}) and AT~2014ej (section  \ref{subsec:fit_AT2014ej}).


\subsection{The cocoon toy model fit of SNhunt120} 
\label{subsec:fit_SNhunt120}
\begin{figure}[ht!]
	\centering
	\includegraphics[trim=3.5cm 8cm 4cm 8cm ,clip, scale=0.62]{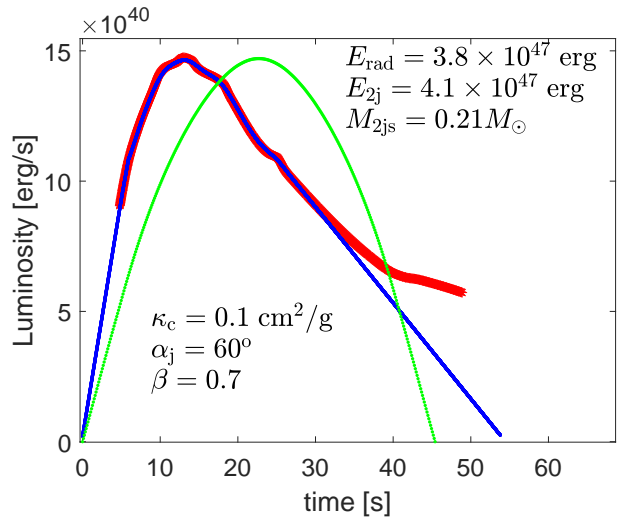}
	\caption{The light curve of SNhunt120 (thick-red line) from \cite{Stritzingeretal2020SNhunt120}, our extension of the peak of light curve (blue line), and a light curve of the cocoon toy model (green line for case 1). 
We constrain the green light curve to fit the total radiated energy of the peak $E_{\rm rad,hunt} = 3.8 \times 10^{47} \erg$ and its maximum luminosity  $L_{\rm hunt} = 1.4 \times 10^{41} \erg \s ^{-1}$. The parameters of this fit (Case 1) are the opacity $\kappa_{\rm c}$, the jets' half opening angle $\alpha_{\rm j}$, and the radius of the jet-shell interaction region  $\beta R_{\rm BB}$, where $R_{\rm BB}$ is the photosphere radius. We calculate the values of the combined energy of the two jets $E_{2{\rm j}}$ and the combined masses in the jets-shell interaction regions $M_{2{\rm js}}$. Note that we do not try to fit the shape of the light curves, but rather only try to explain the amount of radiated energy and maximum luminosity of the peak.
	}
	\label{fig:SNhunt120_LBB_peak}
\end{figure}

First we extend the observed light curve of SNhunt120 (\citealt{Stritzingeretal2020SNhunt120}; thick-red line in Fig. \ref{fig:SNhunt120_LBB_peak}) by taking a linear fit before discovery and beyond $t=30 \days$ after discovery, in both sides down to $L=0$. This is the solid-blue line in Fig. \ref{fig:SNhunt120_LBB_peak}. The observed light curve of this ILOT has a break at about 40 days post-discovery, where the decline becomes shallower. This might result from a second and weaker jet-launching episode or from matter collision in the equatorial plane. We are interested here only in the light curve around the maximum, so we continue the steep decline beyond 30 days post-discovery down to $L=0$. 
We then find the radiated energy of SNhunt120 of our fit to the peak to be $E_{\rm rad,hunt} = 3.8 \times 10^{47} \erg$. As we explained in section \ref{subsec:interaction_TM}, we then build a toy-model symmetric light curve that has the same maximum luminosity as SNhunt120, $L_{\rm hunt} = 1.4 \times 10^{41} \erg \s^{-1}$, and the same total radiated energy. This is the green line in Fig. \ref{fig:SNhunt120_LBB_peak} (for Case 1 that we describe next). 
     
We calculate the energy of one jet $E_{\rm j}$ and the mass in the region of interaction of one jet with the shell (the cocoon), $M_{\rm js}$, as follows. We build a symmetric toy model light curve (one example is the green line in Fig. \ref{fig:SNhunt120_LBB_peak}) that is characterised by its maximum luminosity $L_{\rm j}$ and by its timescale from start to maximum $t_j$ by equations (\ref{eq:jet}). We then calculate the total radiated energy according to this light curve (area under the green light curve). 
We iterate the values of $E_{\rm j}$ and $M_{\rm js}$ until we obtain the luminosity due to the two jets together of $L_{\rm 2j}=L_{\rm hunt} = 1.4 \times 10^{41} \erg \s^{-1}$, and the total radiated energy from the two jets is $E_{\rm rad,2j} = E_{\rm rad,hunt}= 3.8 \times 10^{47} \erg$. 
We note that the cocoon toy model is not sensitive to the expansion velocities of the shell and of the jets, as long as the $v_{\rm j} \gg v_{\rm s}$. 
 
We do not vary the values of the photosphere radius $R_{\rm BB}=2 \times 10^{14} \cm$ that we take from \cite{Stritzingeretal2020SNhunt120}, and of $\beta=0.7$ in equations (\ref{eq:jet}). We do vary the values of the jet's half opening angle $\alpha_j$ and of the opacity $\kappa_{\rm c}$.  
We continue with the wide jets that we discussed in section \ref{sec:Features} \citep{Soker2020ILOT} and scale with $\alpha_{\rm j} = 60^{\circ}$, but we consider narrower jets as well. We scale the opacity with $\kappa_{\rm c}=0.1 \cm^2 \g^{-1}$ but examine also $\kappa_{\rm c}=0.05 \cm^2 \g^{-1}$ and $\kappa_{\rm c}=0.3 \cm^2 \g^{-1}$ to demonstrate the model sensitivity to opacity.
The relevant scaling of equations (\ref{eq:jet}) for SNhunt120, (for one jet) read 
\begin{eqnarray}
\begin{aligned} 
& t_{\rm j} = 22.7    
\left(\frac{E_{\rm j}}{2 \times 10^{47} \erg}\right)^{-1/4} \\ & \times
\left(\frac{M_{\rm js}}{0.1 M_{\odot}}\right)^{3/4}
\left(\frac{\kappa_{\rm c}}{0.1 \cm^2\g^{-1}}\right)^{1/2} {~\rm d}, 
\end{aligned}
\label{eq:tjet_SNhunt120}
\end{eqnarray}
and
\begin{eqnarray}
\begin{aligned} 
& L_{\rm j} = 7.3 \times 10^{40}
\left(\frac{\sin{\alpha_{\rm j}}}{0.87}\right) 
\left(\frac{\beta}{0.7}\right)
\\ & \times
\left(\frac{M_{\rm js}}{0.1 M_{\odot}}\right)^{-3/2}
\left(\frac{E_{\rm j}}{2 \times 10^{47} \erg}\right)^{3/2}
 \\ & \times
\left(\frac{\kappa_{\rm c}}{0.1  \cm^2\g^{-1}}\right)^{-1}
\left(\frac{R_{\rm BB}}{2 \times 10^{14} \cm}\right) \erg \s^{-1}. 
\end{aligned}
\label{eq:Ljet_SNhunt120}
\end{eqnarray}
As with the spherical shell model, we do not assume the energy of the jets. The input variables to the fitting process are the light curve, the half opening angle of the jets, the opacity, and the values of $\beta$ and $\sin \alpha_{\rm j}$. We take the radius of the continuum black body photosphere from observations. We then substitute in equations (\ref{eq:tjet_AT2014ej}) and (\ref{eq:Ljet_AT2014ej}) the observed ILOT's (or one peak of the ILOT) duration $t_{\rm j}$ and the energy radiated from one jet-shell interaction $L_{\rm j}$, and solve for the one jet's energy $E_{\rm j}$ and the mass of the shell that one jet interacts with $M_{\rm js}$.

In Table \ref{Table:SNhunt120_parameters} we present six sets of values in the cocoon toy model for SNhunt120. We emphasise that we do not try to fit the shape of the light curves, and only try to explain the amount of radiated energy, the timescale, and the maximum luminosity of the peak.
In Fig. \ref{fig:SNhunt120_LBB_peak} we show by the green line Case 1.  
\begin{table}[ht!]
\centering
\begin{tabular}{|c|c|c|c|c|c|}
\hline
Case    & $\kappa_{\rm c}$          & $\alpha_{\rm j}$  & $E_{\rm 2j}$          & $M_{\rm 2js}$     & $f_{\rm rad}$      \\
        & (${\rm cm}^2 \g^{-1}$)    &                   & ($10^{47} \erg$)      & ($M_{\odot}$)     &       \\ 
    \hline
    \hline
1           &   $0.1$               & $60^{\circ}$      & $4.1$                 & $0.21$            & $0.93$   \\
    \hline
2           &   $0.3$               & $60^{\circ}$      & $7.2$                 & $0.13$            & $0.53$   \\
    \hline 
3           &   $0.05$              & $40^{\circ}$      & $4.6$                 & $0.36$            & $0.83$   \\
    \hline 
4           &   $0.1$               & $40^{\circ}$      & $6.4$                 & $0.25$            & $0.59$   \\
    \hline
5           &   $0.3$               & $40^{\circ}$      & $11.4$                & $0.15$            & $0.33$   \\
    \hline 
6           &   $0.1$               & $30^{\circ}$      & $9.6$                 & $0.29$            & $0.4$    \\
    \hline 
\end{tabular}
\caption{Six different sets of parameters that fit the peak of the light curve and the total radiated energy of the ILOT SNhunt120 in the frame of the cocoon toy model. The opacity $\kappa_c$ and the jets' half opening angle $\alpha_j$ are input parameters of the modelling. Other parameters are as in equations (\ref{eq:tjet_SNhunt120}) and (\ref{eq:Ljet_SNhunt120}). We calculate from these equations (see text) the combined energy of the two jets $E_{\rm 2j}$ and the combined mass in the interaction regions of the two jets with the shell $M_{\rm 2js}$. In the last column we list the emission efficiency $f_{\rm rad}=E_{\rm rad}/E_{\rm 2j}$.
}
\label{Table:SNhunt120_parameters}
\end{table}

The energy of the jets and the mass they interact with vary between the cases. The energy range is $E_{\rm 2j} \simeq 4\times 10^{47} \erg - 11\times 10^{47} \erg$. In the spherical-shell model of section \ref{subsec:SNhunt120} the jets' energy is $4.5\times 10^{47} \erg$. From the cases of tables \ref{Table:Parameters} and \ref{Table:SNhunt120_parameters} we crudely take the jets' energy for this ILOT to be $E_{\rm 2j}({\rm SNhunt120}) \simeq 5 \times 10^{47} \erg$. 
  
In the cocoon toy model the jets interact with a fraction of the shell. After the `mini-explosion' the assumed spherical interaction zone (cocoon) expands from its initial cocoon-radius $a_{\rm c,0} =\sin \alpha_j \beta R_{\rm BB}$ to larger radii. 
The mass in the interaction zone is then $M_{\rm 2js} > (1-\cos \alpha_j) M_{\rm s}$. Namely, the shell mass is $M_{\rm s} <  M_{\rm 2js}/(1-\cos \alpha_j)$. From Table \ref{Table:SNhunt120_parameters} we find the shell masses of the different cases to be $M_{\rm s} ({\rm Case 2}) < 0.3 M_\odot$ to $M_{\rm s} ({\rm Case 6}) < 2.2 M_\odot$. In the spherical shell model the shell mass is $0.7M_\odot$ (table \ref{Table:Parameters}).
We crudely take for this ILOT $M_{\rm s}({\rm SNhunt120}) \simeq 0.5-1 M_\odot$, but we note that the model can accommodate somewhat lower shell masses.
As we discussed in section \ref{subsec:SNhunt120} the progenitor binary system of this ILOT might have a combined mass of $M_1 + M_2 \approx 10 M_\odot$. 
     
\subsection{The cocoon toy model fit of AT~2014ej} 
\label{subsec:fit_AT2014ej}

\begin{figure}[ht!]
	\centering
	\includegraphics[trim=3.5cm 8cm 4cm 8cm ,clip, scale=0.62]{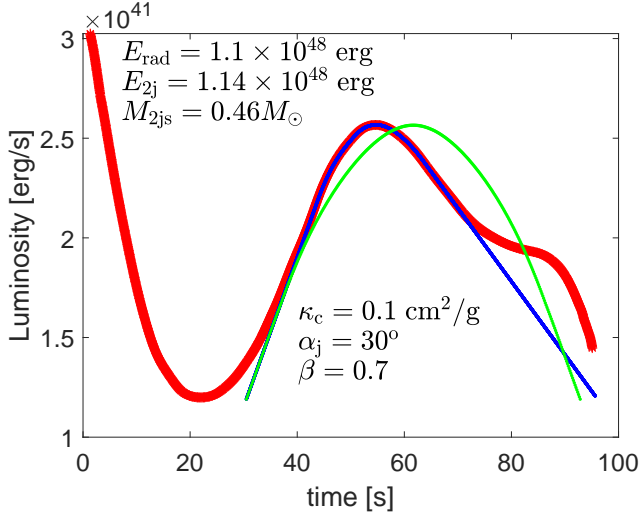}
	\caption{Similar to Fig. \ref{fig:SNhunt120_LBB_peak} but for AT~2014ej. We show the light curve of AT~2014ej (thick-red line; from \citealt{Stritzingeretal2020AT2014ej}), our fit to the light curve of the second peak of AT~2014ej (blue line), and the assumed light curve of the cocoon toy model (green line). We fit the radiated energy of the second peak $E_{\rm rad, AT} = 1.1 \times 10^{48} \erg$ and the maximum luminosity $L_{\rm AT} = 2.6 \times 10^{41} \erg \s ^{-1}$. The relevant scaled-equations are (\ref{eq:tjet_AT2014ej}) and (\ref{eq:Ljet_AT2014ej}).
	}
	\label{fig:AT2014ej_LBB_peak}
\end{figure}

Because at discovery AT~2014ej was already in its decline from the first peak in its light curve, we try to fit the maximum luminosity and the radiated energy of the second peak only. In Fig. \ref{fig:AT2014ej_LBB_peak} we plot by the thick-red line the black-body light curve of AT~2014ej as \cite{Stritzingeretal2020AT2014ej} estimate (their figure 4).
The maximum luminosity of the second peak is $L_{\rm AT} = 2.6 \times 10^{41} \erg \s^{-1}$. In our cocoon toy model this value implies $L_{\rm j} = L_{\rm 2j}/2=L_{\rm AT}/2=1.3 \times 10^{41} \erg \s^{-1}$. 
We examine only the time near maximum luminosity before the break around $t \simeq 70 \days$. We therefore extend the black-body light curve near maximum (solid-blue line in Fig. \ref{fig:AT2014ej_LBB_peak}) by taking a linear fit before $t=42 \days$ and beyond $t=67 \days$ after discovery, in both sides down to $L_{\rm AT}=1.2 \times 10^{41}$, which is the minimum in the light curve between the two peaks. 
We find that the total energy that this ILOT radiated in its second peak according to our fit (solid-blue line in Fig. \ref{fig:AT2014ej_LBB_peak}) is $E_{\rm rad,AT} = 1.1 \times 10^{48} \erg$. We note that in section \ref{subsec:AT2014ej} where we apply the spherical shell model we include the `hump' at $t\simeq90 \days$, and therefore the radiated energy is somewhat larger. The hump can result from a weak third jet-launching episode or from mass collision in the equatorial plane.  
    
We recall that our cocoon toy model does not fit a light curve, but rather fit only the maximum luminosity and total radiated energy (or timescale). We rather assume a symmetric light curve (green line in Fig. \ref{fig:AT2014ej_LBB_peak} for Case 1). We proceed as in section \ref{subsec:fit_SNhunt120} and solve iterativelly equations (\ref{eq:jet}) for several combinations of the input parameters $\alpha_{\rm j}$ and $\kappa_{\rm c}$. We can scale equations (\ref{eq:jet}) with typical values for AT~2014ej (Case 1). The scaled equations read 
\begin{eqnarray}
\begin{aligned} 
& t_{\rm j} = 31    
\left(\frac{E_{\rm j}}{1.14 \times 10^{48} \erg}\right)^{-1/4} \\ & \times
\left(\frac{M_{\rm js}}{0.46 M_{\odot}}\right)^{3/4}
\left(\frac{\kappa_{\rm c}}{0.1 \cm^2\g^{-1}}\right)^{1/2} {~\rm d}, 
\end{aligned}
\label{eq:tjet_AT2014ej}
\end{eqnarray}
and
\begin{eqnarray}
\begin{aligned} 
& L_{\rm j} = 6.9 \times 10^{40}
\left(\frac{\sin{\alpha_{\rm j}}}{0.5}\right) 
\left(\frac{\beta}{0.7}\right)
\\ & \times
\left(\frac{M_{\rm js}}{0.46 M_{\odot}}\right)^{-3/2}
\left(\frac{E_{\rm j}}{1.14 \times 10^{48} \erg}\right)^{3/2}
 \\ & \times
\left(\frac{\kappa_{\rm c}}{0.1  \cm^2\g^{-1}}\right)^{-1}
\left(\frac{R_{\rm BB}}{2.5 \times 10^{14} \cm}\right) \erg \s^{-1}. 
\end{aligned}
\label{eq:Ljet_AT2014ej}
\end{eqnarray}

We examine four cases with different values of $\alpha_{\rm j}$ and $\kappa_{\rm c}$ that we summarise in Table \ref{Table:AT2014ej_parameters}. 
\begin{table}[ht!]
\centering
\begin{tabular}{|c|c|c|c|c|c|}
\hline
Case    & $\kappa_{\rm c}$          & $\alpha_{\rm j}$  & $E_{\rm 2j}$          & $M_{\rm 2js}$     & $f_{\rm rad}$      \\
        & (${\rm cm}^2 \g^{-1}$)    &                   & ($10^{48} \erg$)      & ($M_{\odot}$)     &       \\ 
    \hline
    \hline
1           &   $0.1$               & $30^{\circ}$      & $1.14$                 & $0.46$            & $0.96$   \\
    \hline
2           &   $0.3$               & $30^{\circ}$      & $1.8$                 & $0.24$            & $0.61$   \\
    \hline 
3           &   $0.1$               & $20^{\circ}$      & $1.8$                 & $0.49$            & $0.61$    \\
    \hline 
4           &   $0.3$               & $20^{\circ}$      & $3$                & $0.28$            & $0.37$   \\
    \hline 
\end{tabular}
\caption{Similar to Table \ref{Table:SNhunt120_parameters} but for the second peak of the ILOT AT~2014ej (Fig. \ref{fig:AT2014ej_LBB_peak}), and with the scaling of equations (\ref{eq:tjet_AT2014ej}) and (\ref{eq:Ljet_AT2014ej}). 
}
\label{Table:AT2014ej_parameters}
\end{table}

We find that we can better fit the second peak in the light curve of AT~2014ej with moderately wide jets $\alpha _{\rm j} \simeq 20-30^\circ$. Fitting with wide jets do not give acceptable results. 
For the parameters we list in Table \ref{Table:AT2014ej_parameters} the jets' energies range is $E_{\rm 2j} \simeq 1.14 \times 10^{48} - 3 \times 10^{48} \erg$. In the spherical shell model for the second peak we found this energy to be $1.6 \times 10^{48} \erg$ (Table \ref{Table:Parameters}). 
We take the jets' energy for this ILOT to be $E_{\rm 2j}({\rm AT~2014ej}) \approx 1.5 \times 10^{48} \erg$. For jets' velocity of $ v_{\rm j} = 1000 \km \s^{-1}$ the mass in the jets is then $M_{\rm 2j} \simeq 0.15 M_\odot$. 

We proceed as in section \ref{subsec:fit_SNhunt120} to put an upper limit on the shell mass $M_{\rm s} <  M_{\rm 2js}/(1-\cos \alpha_j)$. 
We calculate from Table \ref{Table:AT2014ej_parameters}   $M_{\rm s} <$ $2 M_\odot$, $1M_\odot$, $3.7M_\odot$, $1.2M_\odot$ for Cases 1, 2, 3, 4 respectively. In the spherical shell model we crudely estimate (Table \ref{Table:Parameters}) the shell mass to be $M_s \approx 1.5 M_\odot$.   
We take the slow shell mass for this ILOT to crudely be $M_{\rm s}({\rm AT~2014ej}) \approx 1-2 M_\odot$. 
If this shell mass holds, then the progenitor binary system of this ILOT cannot be a low mass system, and requires the combined mass to be $M_1 + M_2 > 5 M_\odot$, and more likely $ M_1 + M_2 \ga 10 M_\odot$. 

\section{Summary}
\label{sec:summary}

We apply the jet-power ILOT scenario to two recently studied ILOTs, SNhunt120 \citep{Stritzingeretal2020SNhunt120} and AT~2014ej \citep{Stritzingeretal2020AT2014ej}. In section \ref{sec:Features} we apply the spherical shell model  \citep{Soker2020ILOT}, and in section \ref{sec:BipolarToyModel} we apply the cocoon toy model that we have used to explain some peaks in the light curve of core collapse supernovae \citep{KaplanSoker2020a}. 
In both these models of the jet-power ILOT scenario fast jets catch up with a slower and older shell and collide with it. The collision converts kinetic energy to thermal energy. The post-shock shell and jets gases form a hot bubble, the cocoon. The cocoon cools by photon diffusion that turns to radiation, and by adiabatic expansion. The competition between these processes determine the efficiency of converting kinetic energy, mainly of the jets, to radiation. 

These two models are very crude because we neither conduct hydrodynamic simulations of the interaction nor radiative transfer calculations. As well, we take some parameters to have constant values, in particular the opacity. Even if one does conduct these numerical calculations, the parameter space of the model is very large. Namely, we have no knowledge of the properties of the shell and of the jets, in particular the distribution of the momentum flux of the shell and of the jets with direction and time.  Nonetheless, we did reach our main goal, which is to show that the jet-powered ILOT scenario can account for these two ILOTs. 

We found the following properties of the jet-powered ILOT scenario for these ILOTs. For SNhunt120 (Table \ref{Table:SNhunt120_parameters}) we found that we need to use moderately-wide, $\alpha_j \simeq 30^\circ$, to wide, $\alpha_j \simeq 60^\circ$, jets. 
For wider jets the assumptions of the model break (like that the cocoon has time to expand), and for narrower jets the shell becomes too massive. The typical jets' energy that might explain the peak of SNhunt120 is $E_{\rm 2j}({\rm SNhunt120}) \simeq 5 \times 10^{47} \erg$ (Tables \ref{Table:Parameters} and \ref{Table:SNhunt120_parameters}). 
For jets' velocity of $v_{\rm j} = 1000 \km \s^{-1}$ the mass in the jets is then $M_{\rm 2j} \simeq 0.05 M_\odot$. 
The mass of the shell is less certain, and it is sensitive to the parameters of the models. We crudely estimated $M_{\rm s}({\rm SNhunt120}) \simeq 0.5-1 M_\odot$. 
  
For the second peak of AT~2014ej we had to use moderately wide jets (Table \ref{Table:AT2014ej_parameters}). The jets' energy is $E_{\rm 2j}({\rm AT~2014ej}) \approx 1.5 \times 10^{48} \erg$ (Tables \ref{Table:Parameters} and  \ref{Table:AT2014ej_parameters}). For jets' velocity of $v_{\rm j} = 1000 \km \s^{-1}$ the mass in the jets is then $M_{\rm 2j} \simeq 0.15 M_\odot$. We crudely estimated $M_{\rm s}({\rm AT~2014ej}) \approx 1-2 M_\odot$. 
 
To launch jets with a mass of $\simeq 0.1 M_\odot$ the star that launches the jets should accrete $M_{\rm acc} \simeq 10 M_{\rm 2j} \simeq 1 M_\odot$. An example for such a case is a very young massive star of $\approx 10 M_\odot$ that tidally destroys a pre-main sequence star of $\simeq M_\odot$ and accretes most of its mass. This high value of accreted mass and the massive shell $M_{\rm s} \approx 1M_\odot$, suggest that the binary system progenitors of these two ILOTs are massive, namely $M_1 + M_2 \ga 10 M_\odot$. 
 
Future studies should include more accurate numerical simulations of the jet-shell interaction and of radiative transfer. A parallel line of studies should examine which type of binary systems can lead to such high mass transfer and mass loss rates.  

\section*{Acknowledgements}

We thank Ari Laor for useful discussions and Amit Kashi and an anonymous referee for helpful comments. This research was supported by a grant from the Israel Science Foundation (420/16 and 769/20) and a grant from the Asher Space Research Fund at the Technion.


\label{lastpage}

\begin{thebibliography}{}

\bibitem[Akashi \& Soker(2020)]{AkashiSoker2020} Akashi, M., \& Soker, N.\ 2020, arXiv e-prints, arXiv:2006.01717

\bibitem[Bear et al.(2011)]{Bearetal2011} Bear, E., Kashi, A., \& Soker, N.\ 2011, \mnras, 416, 1965

\bibitem[Berger et al.(2009)]{Berger2009} Berger, E., Soderberg, A. M., Chevalier, R. A., et al. 2009, \apj, 699, 1850

\bibitem[Blagorodnova et al.(2020)]{Blagorodnovaetal2020} Blagorodnova, N., Karambelkar, V., Adams, S.~M., et al.\ 2020, arXiv e-prints, arXiv:2004.04757

\bibitem[Blagorodnova et al.(2017)]{Blagorodnovaetal2017} Blagorodnova, N., Kotak, R., Polshaw, J., et al.\ 2017, \apj, 834, 107

\bibitem[Boian \& Groh(2019)]{BoianGroh2019} Boian, I., \& Groh, J.~H.\ 2019, \aap, 621, A109.

\bibitem[Bond et al.(2003)]{Bondetal2003} Bond, H.~E., Henden, A., Levay, Z.~G., et al.\ 2003, \nat, 422, 405

\bibitem[Cai et al.(2019)]{Caietal2019} Cai, Y.-Z., Pastorello, A., Fraser, M., et al.\ 2019, \aap, 632, L6

\bibitem[Davidson, \& Humphreys(1997)]{DavidsonHumphreys1997} Davidson, K., \& Humphreys, R.~M.\ 1997, \araa, 35, 1

\bibitem[Gilkis et al.(2019)]{Gilkisetal2019} Gilkis, A., Soker, N., \& Kashi, A.\ 2019, \mnras, 482, 4233

\bibitem[Guillochon et al.(2017)]{Guillochonetal2017} Guillochon, J., Parrent, J., Kelley, L.~Z., et al.\ 2017, \apj, 835, 64

\bibitem[Howitt et al.(2020)]{Howittetal2020} Howitt, G., Stevenson, S., Vigna-G{\'o}mez, A., et al.\ 2020, \mnras, 492, 3229

\bibitem[Hubov{\'a}, \& Pejcha(2019)]{HubovaPejcha2019} Hubov{\'a}, D., \& Pejcha, O.\ 2019, \mnras, 489, 891

\bibitem[Ivanova et al.(2013)]{Ivanovaetal2013a} Ivanova, N., Justham, S., Avendano Nandez, J.~L., \& Lombardi, J.~C.\ 2013, Science, 339, 433

\bibitem[Jencson et al.(2019)]{Jencsonetal2019} Jencson, J.~E., Kasliwal, M.~M., Adams, S.~M., et al.\ 2019, \apj, 886, 40

\bibitem[Jones(2020)]{Jones2020} Jones, D.\ 2020, arXiv e-prints, arXiv:2001.03337

\bibitem[Kami{\'n}ski et al.(2015)]{Kaminskietal2015} Kami{\'n}ski, T., Mason, E., Tylenda, R., \& Schmidt, M.~R.\ 2015, \aap, 580, A34

\bibitem[Kaminski et al.(2020)]{Kaminskietal2020Nova1670} Kaminski, T., Menten, K.~M., Tylenda, R., et al.\ 2020, arXiv:2006.10471

\bibitem[Kaminski et al.(2021)]{Kaminskietal2021Nova1670} Kaminski, T., Steffen, W.,  Bujarrabal, V., Tylenda, R.,  Menten, K.~M., \& Hajduk, M.\ 2021,  arXiv:2010.05832

\bibitem[Kaminski et al.(2018)]{Kaminskietal2018} Kaminski, T., Steffen, W., Tylenda, R.,  Young, K.~H., Patel, N.~A., \& Menten, K.~M.\ 2018, \aap, 617, A129

\bibitem[Kaplan \& Soker(2020)]{KaplanSoker2020a} Kaplan, N., \& Soker, N.\ 2020, \mnras, 492, 3013

\bibitem[Kasen \& Woosley(2009)]{KasenWoosley2009} Kasen, D., \& Woosley, S.~E.\ 2009, \apj, 703, 2205.
  
\bibitem[Kashi et al.(2010)]{Kashietal2010} Kashi, A., Frankowski, A., \& Soker, N.\ 2010, \apjl, 709, L11

\bibitem[Kashi et al.(2019)]{Kashietal2019} Kashi, A., Michaelis, A.~M., \& Feigin, L.\ 2019, Galaxies, 8, 2

\bibitem[Kashi \& Soker(2010a)]{KashiSoker2010a} Kashi, A., \& Soker, N.\ 2010a, \apj, 723, 602

\bibitem[Kashi \& Soker(2016)]{KashiSoker2016} Kashi, A., \& Soker, N.\ 2016, Research in Astronomy and Astrophysics, 16, 99

\bibitem[Kashi \& Soker(2017)]{KashiSoker2017} Kashi, A., \& Soker, N.\ 2017, \mnras, 468, 4938

\bibitem[Kasliwal(2011)]{Kasliwal2011} Kasliwal, M.~M.\ 2011, Bulletin of the Astronomical Society of India, 39, 375

\bibitem[Kasliwal et al.(2012)]{Kasliwaletal2012} Kasliwal, M.~M., Kulkarni, S.~R., Gal-Yam, A., et al.\ 2012, \apj, 755, 161

\bibitem[Klencki et al.(2020)]{Klenckietal2020} Klencki, J., Nelemans, G., Istrate, A.~G., \& Chruslinska, M., \ 2020, arXiv e-prints, arXiv:2006.11286

\bibitem[Lipunov et al.(2017)]{Lipunovetal2017} Lipunov, V.~M., Blinnikov, S., Gorbovskoy, E., et al.\ 2017, \mnras, 470, 2339

\bibitem[L{\'o}pez-C{\'a}mara et al.(2020)]{LopezCamaraetal2020} L{\'o}pez-C{\'a}mara, D., Moreno M{\'e}ndez, E., \& De Colle, F.\ 2020, \mnras, 497, 2057


\bibitem[MacLeod \& Loeb(2020)]{MacLeodLoeb2020}  MacLeod, M., \& Loeb, A. 2020, 	arXiv:2003.01123 

\bibitem[MacLeod et al.(2017)]{MacLeodetal2017} MacLeod, M., Macias, P., Ramirez-Ruiz, E., Grindlay, J., Batta, A., \& Montes, G.\ 2017, \apj, 835, 282
   
\bibitem[MacLeod et al.(2018)]{MacLeodetal2018} MacLeod, M., Ostriker, E.~C., \& Stone, J.~M.\ 2018, \apj, 868, 136.

\bibitem[Mason et al.(2010)]{Masonetal2010} Mason, E., Diaz, M., Williams, R.~E., Preston, G., \& Bensby, T.\ 2010, \aap, 516, A108

\bibitem[Mcley \& Soker(2014)]{McleySoker2014} Mcley, L., \& Soker, N.\ 2014, \mnras, 440, 582

\bibitem[Metzger \& Pejcha(2017)]{MetzgerPejcha2017} Metzger, B.~D., \& Pejcha, O.\ 2017, \mnras, 471, 3200
 
\bibitem[Michaelis et al.(2018)]{Michaelisetal2018} Michaelis, A.~M., Kashi, A., \& Kochiashvili, N.\ 2018, \na, 65, 29

\bibitem[Mould et al.(1990)]{Mouldetal1990} Mould, J., Cohen, J., Graham, J.~R., et al.\ 1990, \apjl, 353, L35

\bibitem[Munari et al.(2002)]{Munarietal2002} Munari, U., Henden, A., Kiyota, S., et al.\ 2002, \aap, 389, L51

\bibitem[Muthukrishna et al.(2019)]{MuthukrishnaetalM2019} Muthukrishna, D., Narayan, G., Mandel, K.~S., Biswas, R., \& Hlo{\v z}ek, R.\ 2019, \pasp, 131, 118002
  
\bibitem[Nandez et al.(2014)]{Nandezetal2014} Nandez, J.~L.~A., Ivanova, N., \& Lombardi, J.~C., Jr.\ 2014, \apj, 786, 39

\bibitem[Ofek et al.(2008)]{Ofek2008} Ofek, E.~O., Kulkarni, S.~R., Rau, A., et al.\ 2008, \apj, 674, 447


\bibitem[Pastorello \& Fraser(2019)]{PastorelloFraser2019} Pastorello, A., \& Fraser, M.\ 2019, Nature Astronomy, 3, 676

\bibitem[Pastorello et al.(2018)]{Pastorelloetal2018} Pastorello, A., Kochanek, C.~S., Fraser, M., et al.\ 2018, \mnras, 474, 197

\bibitem[Pastorello et al.(2019)]{PastorelloMasonetal2019} Pastorello, A., Mason, E., Taubenberger, S., et al.\ 2019, \aap, 630, A75

\bibitem[Pejcha et al.(2016a)]{Pejchaetal2016a} Pejcha, O., Metzger, B.~D., \& Tomida, K.\ 2016a, \mnras, 455, 4351
  
\bibitem[Pejcha et al.(2016b)]{Pejchaetal2016b} Pejcha, O., Metzger, B.~D., \& Tomida, K.\ 2016b, \mnras, 461, 2527

\bibitem[Rau et al.(2007)]{Rau2007} Rau, A., Kulkarni, S.~R., Ofek, E.~O., \& Yan, L.\ 2007, \apj, 659, 1536

\bibitem[Retter \& Marom(2003)]{RetterMarom2003} Retter, A., \& Marom, A.\ 2003, \mnras, 345, L25

\bibitem[Schr{\o}der et al.(2020)]{Schrderetal2020} Schr{\o}der, S.~L., MacLeod, M., Loeb, A., Vigna-G{\'o}mez, A., \& Mandel, I.\ 2020, arXiv:1906.04189

\bibitem[Segev et al.(2019)]{Segevetal2019} Segev, R., Sabach, E., \& Soker, N.\ 2019, \apj, 884, 58

\bibitem[Shara et al.(1985)]{Sharaetal1985} Shara, M.~M., Moffat, A.~F.~J., \& Webbink, R.~F.\ 1985, \apj, 294, 271

\bibitem[Soker(2007)]{Soker2007} Soker, N.\ 2007, \apj, 661, 490

\bibitem[Soker(2016)]{Soker2016GEEI} Soker, N.\ 2016, \na, 47, 16

\bibitem[Soker(2020a)]{Soker2020ILOT} Soker, N.\ 2020a, \apj, 893, 20

 
\bibitem[Soker(2020b)]{Soker2020Galaxy} Soker, N.\ 2020b, Galaxies, 8, 26


\bibitem[Soker \& Gilkis(2018)]{SokerGilkis2018} Soker, N., \& Gilkis, A.\ 2018, \mnras, 475, 1198

\bibitem[Soker \& Kashi(2016)]{SokerKashi2016TwoI} Soker, N., \& Kashi, A.\ 2016, \mnras, 462, 217
 
\bibitem[Soker \& Tylenda(2003)]{SokerTylenda2003} Soker, N., \& Tylenda, R.\ 2003, \apjl, 582, L105
  
\bibitem[Stritzinger et al.(2020a)]{Stritzingeretal2020AT2014ej} Stritzinger, M.~D., Taddia, F., Fraser, M., et al.\ 2020a, arXiv e-prints, arXiv:2005.00076


\bibitem[Stritzinger et al.(2020b)]{Stritzingeretal2020SNhunt120} Stritzinger, M.~D., Taddia, F., Fraser, M., et al.\ 2020b, arXiv e-prints, arXiv:2005.00319

\bibitem[Tylenda(2005)]{Tylenda2005} Tylenda, R.\ 2005, \aap, 436, 1009

\bibitem[Tylenda et al.(2011)]{Tylendaetal2011} Tylenda, R., Hajduk, M., Kami{\'n}ski, T., et al.\ 2011, \aap, 528, A114

\bibitem[Tylenda et al.(2013)]{Tylendaetal2013} Tylenda, R., Kami{\'n}ski, T., Udalski, A., et al.\ 2013, \aap, 555, A16
    
\bibitem[Tylenda \& Soker(2006)]{TylendaSoker2006AA} Tylenda, R., \& Soker, N.\ 2006, \aap, 451, 223


\bibitem[Williams et al.(2015)]{Williamsetal2015} Williams, S.~C., Darnley, M.~J., Bode, M.~F., \& Steele, I.~A., \ 2015, \apjl, 805, L18

\bibitem[Yalinewich \& Matzner(2019)]{YalinewichMatzner2019} Yalinewich, A., \& Matzner, C.~D.\ 2019, \mnras, 490, 312

\end{thebibliography}
\end{document}